\documentclass[prl,amsmath, twocolumn]{revtex4}
\usepackage{amsfonts}
\usepackage{pstricks}
\usepackage{pst-poly}
\usepackage{graphicx}
\usepackage{txfonts}
\usepackage{times}
\usepackage{multirow}
\usepackage{amssymb}
\usepackage{times}
\usepackage{CJK}

\newcommand{\beq}{\begin{equation}}
\newcommand{\eeq}{\end{equation}}
\newcommand{\beqa}{\begin{eqnarray}}
\newcommand{\eeqa}{\end{eqnarray}}
\newcommand{\beqar}{\begin{eqnarray*}}
\newcommand{\eeqar}{\end{eqnarray*}}

\def\bi{\begin{itemize}}
\def\ei{\end{itemize}}
\def\be{\begin{equation}}
\def\ee{\end{equation}}
\def\bea{\begin{eqnarray}}
\def\eea{\end{eqnarray}}
\def\ben{\begin{eqnarray*}}
\def\een{\end{eqnarray*}}

\def\>{\rangle}
\def\<{\langle}

\newcommand{\1} I %{{\openone}}

\def\*{\star}

\begin{document}

{\bf Comment on "Quantitative Condition is Necessary in Guaranteeing the Validity of the Adiabatic Approximation"}

Recently, the authors of Ref.\cite{Tong} claimed that they have
proven the traditional adiabatic condition is a necessary condition.
Here, it is claimed that there are some mistakes and an artificial
over-strong constraint in \cite{Tong}, making its result
inconvincible.

In their proof in \cite{Tong}, the author underestimated the
contributions of other small components in a general adiabatic
evolution, which is implied in Eq.(7) in \cite{Tong}. A detailed
version of Eq.(7) in \cite{Tong} is as follows
\begin{equation}
|\psi(t)\rangle=\beta(t)|\psi^{adi}(t)\rangle+\delta(t)|\psi^{adi^\perp}(t)\rangle,
\end{equation}%
where $|\delta(t)|\ll1$ and $\beta(t)^2+|\delta(t)|^2=1$. In the
following, time parameter $t$ is omitted for convenience. Actually,
Eq.(1) is equivalent to Eq.(10.54) in \cite{griffiths}.
Differentiating both sides of Eq.(1) we get
 \begin{equation}
|\dot{\psi}\rangle=\beta|\dot{\psi}^{adi}\rangle+\dot{\beta}|\psi^{adi}\rangle+\dot{\delta}|\psi^{adi^\perp}\rangle+\delta|\dot{\psi}^{adi^\perp}\rangle.
\end{equation}%
Substituting Eq.(1) into the Schrodinger's equation, and combining
the result with Eq.(2), we get($\hbar=1$)
\begin{eqnarray}
i|\dot{\psi}^{adi}\rangle&=&H|\psi^{adi}\rangle-i\frac{\dot{\beta}}{\beta}|\psi^{adi}\rangle+\frac{1}{\beta}\big(H\delta|\psi^{adi^\perp}\rangle\notag\\&&-i\dot{\delta}|\psi^{adi^\perp}\rangle-i\delta|\dot{\psi}^{adi^\perp}\rangle\big).
\end{eqnarray}%
As $\dot{\beta}=-\frac{|\delta|}{\beta}\frac{d|\delta|}{dt}$, a term
which contains $\dot{\beta}$ is subleading in the right side of
Eq.(3). To simplify the following discussion we may neglect its
effect. So Eq.(6) in \cite{Tong} declared to be a necessary
condition is valid if and only if
$\left\|\frac{1}{\beta}(-i\dot{\delta}|\psi^{adi^\perp}\rangle-i\delta|\dot{\psi}^{adi^\perp}\rangle
)\right\|$  is small comparing to
$\left\|H|\psi^{adi}\rangle\right\|$. Here and thereafter the symbol
$\|\cdot\|$ denotes the norm of vector. This generally requires
$\left|\dot{\delta}\right|\ll E_m$ and
$\left\|\delta|\dot{\psi}^{adi^\perp}\rangle\right\|\ll E_m$ which
are extra requirements in addition to the existing requirement of
Eq.(7) in \cite{Tong}.

It's explicit that the 'Proof' in \cite{Tong} can not be completed
without the extra limitation Eq.(6) in \cite{Tong}. Substituting
Eq.(1) and Eq.(2) into Eq.(12) in \cite{Tong} and neglect the term
containing $\dot{\beta}$, we get
\begin{eqnarray}
c_m=\frac{1}{E_m-E_n}[i\beta
\gamma_1+(i\dot{\delta}-E_n\delta)\gamma_2 +i\delta\gamma_3)],
\end{eqnarray}
where $\gamma_1=\langle
E_m|\dot{\psi}^{adi}\rangle$,$\gamma_2=\langle
E_m|\psi^{adi^\perp}\rangle$,$\gamma_3=\langle
E_m|\dot{\psi}^{adi^\perp}\rangle$. Terms containing $\delta$ can be
ignored assuming $\delta$ is small enough,
\begin{eqnarray}
c_m=\frac{1}{E_m-E_n}(i\beta \gamma_1+i\dot{\delta}\gamma_2),
\end{eqnarray}
but terms containing $\dot{\delta}$ cannot be ignored
\cite{griffiths}. Comparing Eq.(5) here with Eq.(13) in \cite{Tong},
one can see that the proof in \cite{Tong} can only be applied to
special systems. It is quite possible that neither
 of the two terms in the right side of Eq.(5) is small but
their summation is small. In this case, quantitative condition is
unnecessary. Specifically, the second example in section V of
\cite{Jianda} supports the conclusion.

Moreover, there is a loophole in the logic of the "proof" in
\cite{Tong}. $E_n$ in its Eq.(11) can be replaced by any number
other than $E_m$ to complete its following "proof". If a replacement
is done, then nothing useful can be deduced by its "proof". The
loophole arises from the suspicious $"\simeq"$ in Eq.(13) in
\cite{Tong}. In fact, the second term in the right side of Eq.(12)
in \cite{Tong}, $\frac{\langle
E_m|E_n|\psi\rangle}{E_m-E_n}=\frac{E_n c_m}{E_m-E_n}$, is at the
same order of (or even much larger than) $c_m$. It can not be simply
substituted by $\frac{\langle
E_m|E_n|\psi^{adi}\rangle}{E_m-E_n}=0$.

In conclusion, Ref. \cite{Tong} accomplished its 'proof' by
introducing an extra constraint, Eq.(6) in \cite{Tong}, which means
their "necessary condition" can only be applied to a small class of
systems. In addition there is a logical loophole in its proof.
Thereby the proof is generally not reliable and their comments on
others' models in other papers are unreasonable.

One direct reason why adiabatic approximation is important is it can
provide an approximate but convenient approach to control
complicated evolution in quantum systems. And the geometric phase
developed in 1980's further widens its applications. Therefore we
should only be concerned with the fidelity between the adiabatic
state and evolving state, a simple and natural choice. Some special
models\cite{Du} may be noted as models failing to achieve geometric
phase. However, it's not proper to exclude them from being classified 
as adiabatic approximation processes with an extra artificial requirement
(see Eq.(6) in \cite{Tong}). We believe a general necessary
condition should be deduced based on the only requirement that the
adiabatic state should be close to the evolution state. 

Meisheng Zhao$^{1 \dag}$ and Jianda Wu$^{2 \star}$

$^1$Department of Modern Physics, University of Science and
Technology of China, Hefei 230026, People's Republic of China;
$^2$Department of Physics $\&$ Astronomy, Rice University, Houston,
Texas 77005, USA


$^\dag$zmesson@mail.ustc.edu.cn

$^\star$jw5@rice.edu

\end{document}